\documentclass[11pt]{article}
\usepackage[utf8]{inputenc}
\usepackage[T1]{fontenc}
\usepackage{lmodern}
\usepackage{amsmath,amssymb}
\usepackage{graphicx}
\usepackage[margin=1in]{geometry}
\usepackage{xcolor}
\usepackage{float}
\usepackage[colorlinks=true,linkcolor=blue,citecolor=blue,urlcolor=blue]{hyperref}

\newif\ifmarked
\markedfalse

\newcommand{\vv}[1]{\vec{#1}}
\newcommand{\eps}{\varepsilon_0}

\title{Light propagation in dielectric materials:\\ a microscopic derivation of the refractive index\\ and of the extinction theorem}
\author{Giancarlo Reali$^{1}$\thanks{Corresponding author: reali@unipv.it} \ and Ledo Stefanini$^{2}$\\[6pt]
\small $^1$Dipartimento di Ingegneria Industriale e dell'Informazione, Universit\`a di Pavia,\\
\small via Ferrata 5a, 27100 Pavia, Italy\\[3pt]
\small $^2$Accademia Nazionale Virgiliana, Via dell'Accademia 47, 46100 Mantova, Italy}
\date{}

\begin{document}
\maketitle

\begin{abstract}
We present a compact, self-contained derivation of the field established inside a dielectric by an incident light wave, starting from the picture of the material as a collection of polarizable atoms (or molecules) in vacuum, each driven by the incident field and by the fields re-radiated by all the others. Following the integral-equation route of Goldberger and Watson, the same few lines of calculation yield the Lorentz local field, the Lorentz--Lorenz relation between the refractive index and the atomic polarizability, and the Ewald--Oseen extinction theorem: the incident wave is cancelled inside the medium by a field radiated by the induced dipoles and replaced by a wave travelling at $c/n$. The dilute limit is connected to the forward-scattering amplitude, the optical theorem and Rayleigh scattering. We then discuss, for a planar interface, in what sense reflection and refraction are a bulk effect for s-polarized light and a combination of bulk and surface (bound-charge) effects for p-polarized light, and why that surface term must not be confused with the local-field correction, although both stem from the same depolarization mathematics. Finally, a small coupled-dipole simulation of a random, finite collection of point dipoles checks the self-consistent local-field relation numerically. SI units are used throughout. The material is intended for an advanced undergraduate course on electromagnetism and optics.
\end{abstract}

\section{Introduction}

That light is slowed down in glass is usually derived in one line from the macroscopic Maxwell equations, once the polarization is expressed through the refractive index. The derivation is efficient, but it hides the physical mechanism. Glass is mostly empty space; the incident wave and the waves re-radiated by the atoms all travel at $c$ in the interatomic vacuum. Where does the incident wave go, and how does a wave travelling at $c/n$ arise?

The answer, the Ewald--Oseen extinction theorem~\cite{born_wolf}, has a substantial didactic literature. Classic treatments integrate the dipolar fields over a semi-infinite medium or a slab~\cite{born_wolf,reali1982,fearn}; others treat the medium as vacuum with explicit sources~\cite{hagenvik,lai_lau_wong}, discuss the extinction length~\cite{ballenegger,berman}, or address directly whether light ``really'' slows down~\cite{jakoby}. Feynman's \emph{Lectures} give a famously economical account, based on the field radiated by a sheet of oscillating charges~\cite{feynman_31,feynman_33}, which one of us has worked out in detail for reflection at a glass surface~\cite{reali2014}.

The purpose of this paper is not to add a new result to that literature but to offer a short route through it, suitable for two lectures of an advanced course. We adapt to classical electrodynamics the treatment given by Goldberger and Watson in their monograph on collision theory~\cite{goldberger_watson}, which is little known to physics teachers and has a distinctive merit: a single integral equation, solved in a few lines, delivers at once the local field, the Lorentz--Lorenz relation and the extinction theorem, with no assumption on the shape of the sample or on the direction of incidence. We then use this picture for three purposes: (i) to connect the dilute limit to the forward-scattering amplitude and the optical theorem; (ii) to clarify the respective roles of the bulk polarization current and of the surface bound charge in reflection, and to separate the latter from the local-field correction; (iii) to show, with a modest numerical experiment that students can reproduce, that the continuum argument is already realized by a few hundred randomly placed point dipoles. The approximations used are stated, with their physical motivation, in Section~\ref{sec:model}.

\section{The macroscopic description}
\label{sec:maxwell}

In a homogeneous, isotropic, non-magnetic and linear dielectric, the polarization current $\vv J_P=\partial\vv P/\partial t$ acts as a source in the wave equation for the electric field:
\begin{equation}
\nabla^2\vv E-\frac{1}{c^2}\frac{\partial^2\vv E}{\partial t^2}=\mu_0\frac{\partial^2\vv P}{\partial t^2}.
\label{eq:wave}
\end{equation}
With the constitutive relation
\begin{equation}
\vv P=\eps\chi\vv E=\eps(n^2-1)\vv E ,
\label{eq:constitutive}
\end{equation}
and $\mu_0\eps=1/c^2$, Eq.~\eqref{eq:wave} becomes
\begin{equation}
\nabla^2\vv E-\frac{1}{c^2}\frac{\partial^2\vv E}{\partial t^2}=\frac{n^2-1}{c^2}\frac{\partial^2\vv E}{\partial t^2}.
\label{eq:wave_n}
\end{equation}
The source term already contains, in compressed form, the whole story told below: one part of it cancels the vacuum propagation term on the left-hand side (extinction), the rest turns the equation into a homogeneous wave equation with phase velocity $c/n$.

\section{The microscopic model}
\label{sec:model}

\subsection{Assumptions}

We describe the dielectric as $N$ identical polarizable units, randomly distributed with uniform average density $\rho=N/V$ inside a volume $V$ whose boundary has radii of curvature large compared with the wavelength $\lambda$. The units may be atoms (noble gases) or molecules; for molecules, $\alpha$ below is the orientation-averaged polarizability, so that the medium is isotropic. The following assumptions are used, each for a definite purpose:
\begin{itemize}
\item \emph{Dipole approximation}: $\lambda$ is much larger than the size $a$ of the units ($\lambda/a\approx10^3$--$10^4$ for visible light), so each unit radiates as a point dipole.
\item \emph{Frequency below the lowest absorption resonance}: the scattering is elastic and the response linear, described by a polarizability $\alpha(\omega)$.
\item \emph{Impenetrable units}: two units never overlap. Physically obvious, this assumption is what makes it legitimate to exclude from the sum over the neighbours of a given unit a small sphere around it, Eq.~\eqref{eq:En_cont} below. It will reappear, concretely, in the simulation of Section~\ref{sec:numerics}.
\item \emph{Random (or cubic) arrangement}: the near neighbours of a unit are distributed with no preferred direction, so that their quasi-static fields average to zero at its position. This is the hypothesis behind the Lorentz local field; it fails for strongly correlated non-cubic structures and for polar molecules, where the Lorentz--Lorenz relation is known to be only approximate.
\end{itemize}
No assumption on density is needed: the same equations cover dilute gases and dense media, the dilute case being simply the limit $\rho\alpha/\eps\ll1$ (Section~\ref{sec:dilute}).

\subsection{The multiple-scattering equations}

For a monochromatic field ($e^{-i\omega t}$ understood, $k=\omega/c$) the field acting on unit $n$ at $\vv r_n$ is the incident field plus the fields radiated by all the other units:
\begin{equation}
\vv E_n(\vv r_n)=\vv E_0(\vv r_n)+\sum_{m\neq n}\vv E_{\mathrm{rad},m}(\vv r_n).
\label{eq:multiple}
\end{equation}
The structure is self-consistent, since each $\vv E_{\mathrm{rad},m}$ depends on the field acting on unit $m$, given by an equation of the same form.

The field of an oscillating point dipole $\vv p_m$ follows from its retarded Hertz vector,
\begin{equation*}
\vv\Pi=\frac{\vv p_m}{4\pi\eps}\frac{e^{ikr_{nm}}}{r_{nm}},\qquad \vv E=\nabla_n\times\nabla_n\times\vv\Pi ,
\end{equation*}
valid outside the source~\cite{jackson}; with $\vv r_{nm}=\vv r_n-\vv r_m$ and $\nabla(e^{ikr}/r)=-(\vv r/r^2)(1/r-ik)e^{ikr}$ this gives
\begin{equation}
\vv E_{\mathrm{rad},m}(\vv r_n)=\frac{1}{4\pi\eps}\,\nabla_n\times\left[\left(\vv p_m\times\frac{\vv r_{nm}}{r_{nm}^2}\right)\left(\frac{1}{r_{nm}}-ik\right)e^{ikr_{nm}}\right],
\label{eq:dipole}
\end{equation}
which contains both the near (quasi-static, $1/r^3$) and the far (radiation, $1/r$) fields; $\nabla_n$ acts on $\vv r_n$ only. The induced dipole moment is
\begin{equation}
\vv p_m=\alpha\,\vv E_m(\vv r_m),
\label{eq:p}
\end{equation}
and, for the electron of charge $e$ and mass $m_e$ in the standard Lorentz model (one resonance $\omega_0$, damping $\gamma$),
\begin{equation}
\alpha=\alpha'+i\alpha''=\frac{e^2/m_e}{\omega_0^2-\omega^2-i\gamma\omega}.
\label{eq:alpha}
\end{equation}
Equation~\eqref{eq:alpha} is textbook material~\cite{feynman_31,nussenzveig}; nothing below depends on its detailed form except the Rayleigh cross section of Section~\ref{sec:dilute}.

\subsection{Local field and macroscopic field}

Passing to a continuum description, we replace the sum in Eq.~\eqref{eq:multiple} with an integral over $V$, from which a sphere $S_0$ of atomic radius centred on $\vv r_n$ is excluded. Thanks to impenetrability, this excludes unit $n$ and nothing else:
\begin{equation}
\vv E_n(\vv r_n)=\vv E_0(\vv r_n)+\rho\int_{V-S_0}d^3r_m\,\vv E_{\mathrm{rad},m}(\vv r_n).
\label{eq:En_cont}
\end{equation}
Together with Eqs.~\eqref{eq:dipole} and~\eqref{eq:p}, this is an integral equation for the local field. Following Lorentz, the \emph{macroscopic} field is defined by excluding instead an infinitesimally thin needle-shaped cylinder through $\vv r$, parallel to the field:
\begin{equation}
\vv E(\vv r)=\vv E_0(\vv r)+\rho\int_{V-\mathrm{cyl}}d^3r_m\,\vv E_{\mathrm{rad},m}(\vv r).
\label{eq:Emacro}
\end{equation}
The needle is the exclusion region that removes the self-contribution without introducing any depolarization of its own (Appendix~\ref{app:depol}); this is what makes Eq.~\eqref{eq:Emacro} the field of the macroscopic Maxwell equations. Subtracting Eq.~\eqref{eq:Emacro} from Eq.~\eqref{eq:En_cont} at $\vv r=\vv r_n$,
\begin{equation}
\vv E_n(\vv r_n)=\vv E(\vv r_n)-\rho\int_{S_0-\mathrm{cyl}}d^3r_m\,\vv E_{\mathrm{rad},m}(\vv r_n),
\label{eq:difference}
\end{equation}
where the remaining integral extends only over the atomic-size region between the sphere and the needle. There $kr_{nm}\ll1$, so that Eq.~\eqref{eq:dipole} reduces to its quasi-static form, and the dipole density can be taken as uniform, $\vv p_m\simeq\vv p_n$. Exchanging the roles of $n$ and $m$ (Appendix~\ref{app:integral}),
\begin{equation}
\vv E_{\mathrm{rad},m}(\vv r_n)\simeq\frac{1}{4\pi\eps}\,\nabla_n\times\left(\vv p_m\times\frac{\vv r_{nm}}{r_{nm}^3}\right)\;\longrightarrow\;\frac{1}{4\pi\eps}\,\nabla_m\!\left(\frac{\vv p_n\cdot\vv r_{nm}}{r_{nm}^3}\right),
\label{eq:quasistatic}
\end{equation}
the arrow indicating that only the gradient term survives the integration: this is just the field of a static dipole written as the gradient of its potential. Gauss's theorem turns the volume integral into a surface integral, which vanishes on the needle and gives on the sphere
\begin{equation}
\int_{S_0-\mathrm{cyl}}d^3r_m\,\vv E_{\mathrm{rad},m}(\vv r_n)=-\frac{\vv p_n}{3\eps}.
\label{eq:sphere}
\end{equation}
With $\vv P=\rho\vv p_n$, Eq.~\eqref{eq:difference} becomes the Lorentz local field,
\begin{equation}
\vv E_n=\vv E+\frac{\vv P}{3\eps}.
\label{eq:Elocal}
\end{equation}

\subsection{Extinction theorem and Lorentz--Lorenz relation}

Multiplying Eq.~\eqref{eq:En_cont} by $\rho\alpha$, and using Eq.~\eqref{eq:difference} for the integral, we obtain
\begin{equation}
\vv P(\vv r)=\rho\alpha\,\vv E_0(\vv r)+\rho\alpha\left[\vv E(\vv r)-\vv E_0(\vv r)+\frac{\vv P(\vv r)}{3\eps}\right].
\label{eq:selfconsistent}
\end{equation}
This is the self-consistent solution of the integral equation~\eqref{eq:En_cont}. It holds for any direction of $\vv E_0$ and of $\vv k$, and for any sample shape compatible with the assumptions. The polarization it determines is the source of everything that propagates inside the dielectric or emerges from it: the internal and scattered fields and, for planar interfaces, the reflected, refracted and transmitted waves.

Equation~\eqref{eq:selfconsistent} shows at a glance the two effects of the re-radiated fields (the square bracket). First, they contain $-\vv E_0$: inside the medium the incident wave is cancelled by a field radiated by the induced dipoles. This is the \emph{extinction theorem}. Second, they contain $\vv P/3\eps$: the polarization reappears on the right-hand side, the signature of multiple scattering. After the cancellation, with $\vv P\parallel\vv E$ in an isotropic medium and Eq.~\eqref{eq:constitutive},
\begin{equation}
\vv P=\frac{\rho\alpha}{1-\rho\alpha/3\eps}\,\vv E=\eps(n^2-1)\vv E,
\qquad\text{i.e.}\qquad
\frac{n^2-1}{n^2+2}=\frac{\rho\alpha}{3\eps},
\label{eq:LL}
\end{equation}
which is the Lorentz--Lorenz (or Clausius--Mossotti) relation between the microscopic parameters and the refractive index. The internal field obeys Eq.~\eqref{eq:wave_n} and therefore travels at $c/n$, although all the fields that compose it travel at $c$.

\subsection{The dilute limit: forward scattering and the optical theorem}
\label{sec:dilute}

For a gas, $\rho\alpha/\eps\ll1$ and $n-1\ll1$, and Eq.~\eqref{eq:LL} reduces to
\begin{equation}
n=1+\frac{\rho\alpha}{2\eps},
\label{eq:dilute}
\end{equation}
the local-field enhancement being negligible. The same result can be obtained from scattering theory, which illuminates its meaning. In the radiation zone, Eq.~\eqref{eq:dipole} gives for the field scattered in the forward direction by a unit driven by a field of unit amplitude $f(0)\,e^{ikr}/r$, with forward-scattering amplitude
\begin{equation}
f(0)=\frac{k^2\alpha}{4\pi\eps}.
\label{eq:f0}
\end{equation}
Consider now a thin, wide sheet of gas of thickness $L$ in the plane $z=0$, illuminated at normal incidence by a plane wave of unit amplitude. Summing the forward-scattered wavelets over the Fresnel zones of the sheet~\cite{feynman_31,nussenzveig} gives, at $z\gg L$,
\begin{equation}
E(z)=e^{ikz}\left[1+\frac{2\pi i\rho L}{k}f(0)\right]\simeq e^{ikz}\,e^{i2\pi\rho Lf(0)/k}.
\label{eq:sheet}
\end{equation}
The extra phase must equal $(n-1)kL$, whence
\begin{equation}
n=1+\frac{2\pi\rho}{k^2}f(0),
\label{eq:n_f0}
\end{equation}
which, with Eq.~\eqref{eq:f0}, coincides with Eq.~\eqref{eq:dilute}. The factor $i$ in Eq.~\eqref{eq:sheet} is the famous $90^\circ$ lag of the field radiated by a sheet of dipoles, which Feynman uses to explain the phase delay of light in glass~\cite{feynman_31}.

If $f(0)$ has an imaginary part, the intensity transmitted forward decreases,
\begin{equation}
|E(z)|^2=e^{-4\pi\rho L\,\mathrm{Im}f(0)/k}\equiv e^{-\rho\sigma L},
\end{equation}
and, since in our model the only loss is scattering out of the beam, $\sigma$ is the scattering cross section. This is the optical theorem,
\begin{equation}
\sigma=\frac{4\pi}{k}\,\mathrm{Im}f(0)=\frac{k\alpha''}{\eps},
\label{eq:optical}
\end{equation}
here seen as the result of the interference between the incident wave and the forward-scattered one. With Eq.~\eqref{eq:alpha} and the radiative damping constant $\gamma=e^2\omega^2/(6\pi\eps m_ec^3)$~\cite{jackson}, Eq.~\eqref{eq:optical} becomes
\begin{equation}
\sigma=\frac{8\pi}{3}\,r_e^2\,\frac{\omega^4}{(\omega_0^2-\omega^2)^2+\gamma^2\omega^2},
\qquad r_e=\frac{e^2}{4\pi\eps m_ec^2},
\label{eq:rayleigh}
\end{equation}
which for $\omega\ll\omega_0$ is Rayleigh's $\omega^4$ law, responsible for the blue of the sky.

\section{Reflection: bulk and surface contributions}
\label{sec:reflection}

\begin{figure}[!htbp]
\centering
\includegraphics[width=0.4\textwidth]{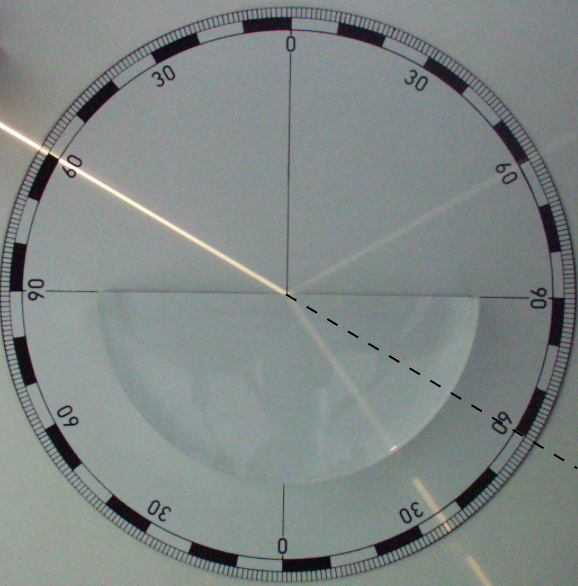}
\caption{A light beam striking obliquely an air--glass interface. The dashed line is the straight continuation of the incident beam inside the glass, which is not observed.}
\label{fig:experiment}
\end{figure}

Figure~\ref{fig:experiment} shows the familiar classroom demonstration of reflection and refraction at an air--glass interface. Feynman reads it backwards~\cite{feynman_33}: no beam is observed along the dashed line, the straight continuation of the incident beam, which would be there if the atoms of the glass could not move. Since the atoms do oscillate, the induced polarization current must radiate, besides the refracted beam, a beam travelling at $c$ that exactly cancels the incident one inside the glass. This is the extinction theorem, used without naming it.

The standard derivation of the Fresnel coefficients imposes boundary conditions on the macroscopic fields at the interface, and the reflected wave then seems to be produced at the surface. The microscopic picture says otherwise: all fields except the incident one are radiated by the polarization established throughout the volume, which, for a planar boundary, organizes itself coherently into the extinction, refracted and reflected waves. The two polarizations must, however, be distinguished. For s (TE) polarization, $\vv P$ is parallel to the interface and the reflected and transmitted fields are produced entirely by the volume polarization current. For p (TM) polarization, the normal component of $\vv P$ is discontinuous at the interface, and the corresponding bound surface charge $\sigma_b=\vv P\cdot\hat n$ radiates an additional contribution, of the same order as the volume one and not reducible to it. This is made explicit by the integral equations of Lai, Lau and Wong for a slab~\cite{lai_lau_wong}: the TE equation contains only a volume term, the TM equation also a term involving only the polarization charges on the two faces. The surface term is no mystery: it is the bound charge implicit in $\vv P=\eps\chi\vv E$, made visible where $\vv P$ is discontinuous. The polarization-dependent angular factors of the reflected and extinction fields are worked out, along Feynman's lines, in Ref.~\cite{reali2014}.

It is tempting to identify this surface term with the local-field correction of Eq.~\eqref{eq:Elocal}, since both come from integrating a dipolar kernel over an anisotropic region and both are depolarization effects (Appendix~\ref{app:depol}). The identification would be wrong. The local-field correction is intrinsic and microscopic: it exists in an unbounded medium and stems from the discreteness of matter, i.e.\ from the small sphere excluded around each unit. The surface term is extrinsic and macroscopic: it exists only because the medium ends, and would be present unchanged in a hypothetical continuous dielectric. The first answers the question ``what field does a unit feel inside its own medium?'', the second ``what does a finite piece of that medium radiate from its boundary?''. They share a mathematical technique, not a physical origin.

\section{A numerical check with coupled dipoles}
\label{sec:numerics}

The passage from the sum~\eqref{eq:multiple} to the integral~\eqref{eq:En_cont} is the key step of Section~\ref{sec:model}. It can be checked directly by solving the discrete problem for a finite random set of point dipoles, the ``coupled-dipole method''~\cite{purcell_pennypacker,draine_flatau}, and asking whether the result follows the Lorentz--Lorenz relation~\eqref{eq:LL} rather than the dilute formula~\eqref{eq:dilute}. The check is meant to be pedagogical and self-contained, not a high-precision computation.

\paragraph{Method.} $N=600$ point dipoles are placed at random in a thin slab $0<z<L$, illuminated by $\vv E_0=\hat x\,e^{ikz}$. Equation~\eqref{eq:dipole} is used in its explicit form~\cite{jackson}
\begin{equation}
\vv E_{\mathrm{rad}}(\vv r)=\frac{1}{4\pi\eps}\,\mathsf G(\vv r)\cdot\vv p,\qquad
G_{ij}=k^2(\delta_{ij}-\hat n_i\hat n_j)\frac{e^{ikr}}{r}+(3\hat n_i\hat n_j-\delta_{ij})\left(\frac{1}{r^3}-\frac{ik}{r^2}\right)e^{ikr},
\label{eq:G}
\end{equation}
with $\hat n=\vv r/r$, retaining full retardation. Equations~\eqref{eq:multiple} and~\eqref{eq:p} become the $3N\times3N$ complex linear system
\begin{equation}
\vv p_m-\frac{\alpha}{4\pi\eps}\sum_{n\neq m}\mathsf G(\vv r_m-\vv r_n)\cdot\vv p_n=\alpha\,\vv E_0(\vv r_m),\qquad m=1,\dots,N,
\label{eq:system}
\end{equation}
which is solved exactly. The polarization $P=\rho\langle p_x\rangle$ is averaged over the dipoles in a thin layer about the mid-plane; the macroscopic field $E$ is computed at the same depth from Eq.~\eqref{eq:Emacro}, i.e.\ excluding a thin needle along $\hat x$ (a spherical exclusion would instead give the local field~\eqref{eq:Elocal}). The index is then read from Eq.~\eqref{eq:constitutive}, $n=\sqrt{1+P/\eps E}$. Since $P$ and $E$ are taken at the same depth, this tests the local relation~\eqref{eq:LL}, not the phase accumulated over a propagation distance. The only material parameter is the dimensionless combination
\begin{equation*}
x=\frac{\rho\alpha}{4\pi\eps},
\end{equation*}
in terms of which Eq.~\eqref{eq:dilute} reads $n=1+2\pi x$ and Eq.~\eqref{eq:LL} reads $(n^2-1)/(n^2+2)=4\pi x/3$. We used $\lambda=550$\,nm, a slab $L=0.3/k\approx26$\,nm thick with transverse size $(1/k)^2$, $1/k\approx87.5$\,nm. The resulting spacing between dipoles, about $8$\,nm, is much larger than in a real solid; this coarse graining only changes the values of $\rho$ and $\alpha$ needed for a given $x$.

\begin{figure}[!htbp]
\centering
\includegraphics[width=0.66\textwidth]{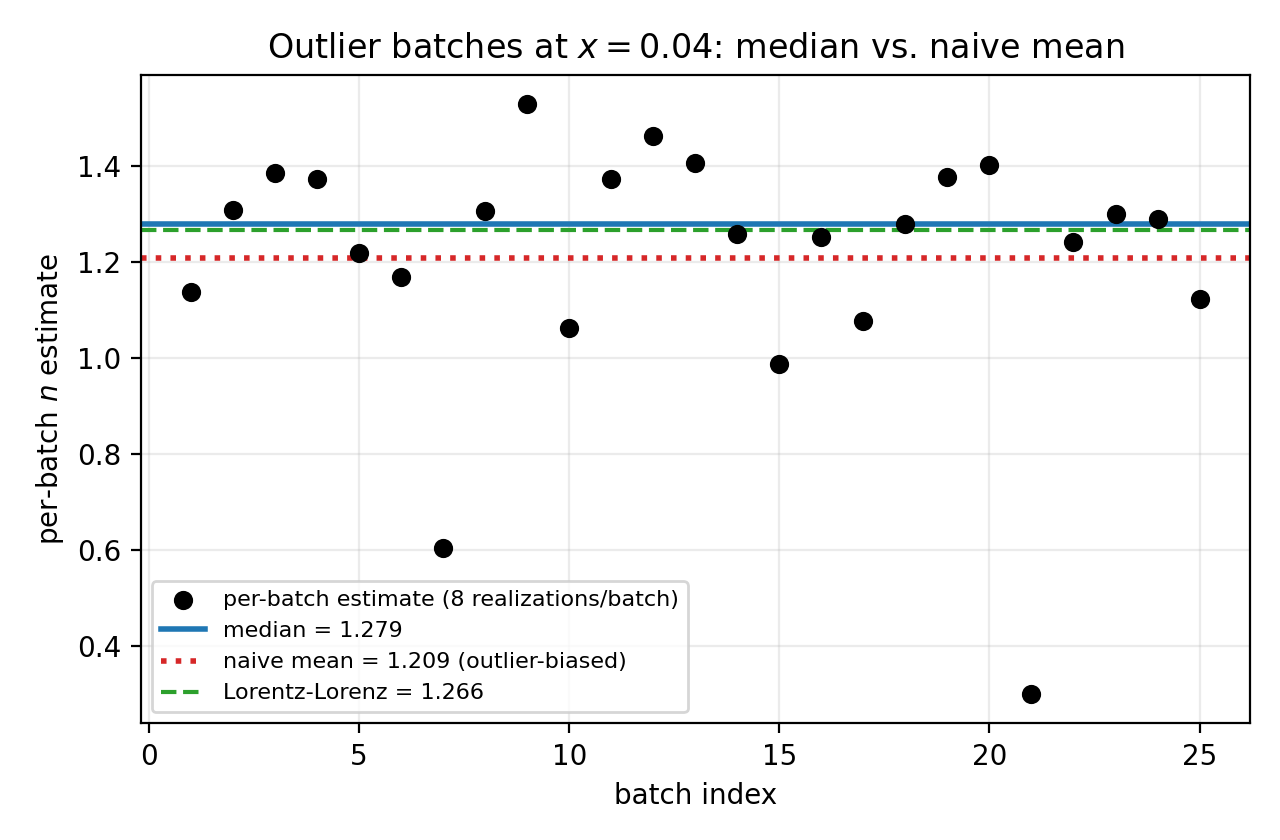}
\caption{Estimates of $n$ from 25 independent batches (8 realizations each, $N=600$) at $x=0.04$. Two batches are pulled far down by rare strongly coupled configurations: the median (solid) is robust, the mean (dotted) is biased. Dashed: Lorentz--Lorenz value.}
\label{fig:outlier}
\end{figure}

\paragraph{Two numerical lessons.} Treating the units as mathematical points creates two difficulties, both echoes of the physics. First, random positions occasionally put two dipoles very close; the $1/r^3$ near field then produces unphysically large moments. Impenetrability must be restored by hand, rejecting configurations with pairs closer than a hard-core radius: this is the discrete counterpart of the excluded sphere $S_0$. Second, even with a hard core, a few realizations contain clusters of strongly coupled neighbours that produce large outliers (Fig.~\ref{fig:outlier}). A batch median, with the median absolute deviation as error estimate, is far more robust than the mean. The effect grows with density: it is a small-scale precursor of the Clausius--Mossotti singularity at $\rho\alpha/3\eps\to1$.

\begin{figure}[!htbp]
\centering
\includegraphics[width=0.66\textwidth]{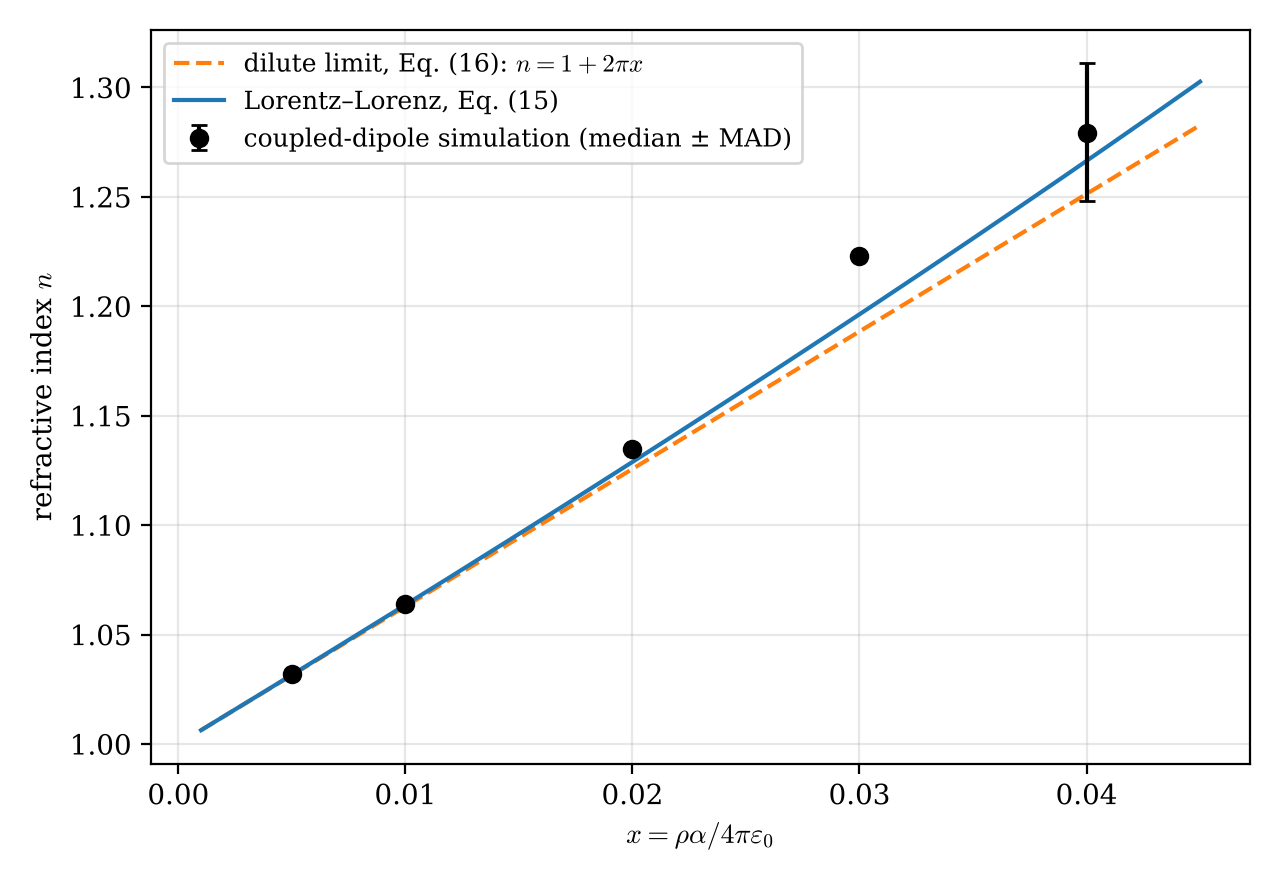}
\caption{Refractive index from the coupled-dipole simulation (points: median $\pm$ MAD over 15--25 batches; error bars smaller than the symbols except at $x=0.04$), compared with the dilute formula, Eq.~\eqref{eq:dilute} (dashed), and the Lorentz--Lorenz relation, Eq.~\eqref{eq:LL} (solid).}
\label{fig:nvsx}
\end{figure}

\paragraph{Results.} Figure~\ref{fig:nvsx} compares the simulated $n$ with Eqs.~\eqref{eq:LL} and~\eqref{eq:dilute}. At low density ($x\lesssim0.01$), where the two formulas nearly coincide, the agreement is at the sub-percent level. At higher density the data follow the Lorentz--Lorenz curve, with a systematic offset at the percent level, probably due to the finite size of the sample, and a growing statistical uncertainty; beyond $x\approx0.04$ the outliers make the estimate too noisy to be useful without much larger computational effort. Within these limits, the simulation shows that the continuum argument of Section~\ref{sec:model} is already realized by a few hundred randomly placed impenetrable dipoles.

\section{Conclusion}

A single integral equation for the field acting on each atom, solved with Lorentz's distinction between local and macroscopic field, yields in a few lines the extinction of the incident wave, its replacement by a wave travelling at $c/n$, and the Lorentz--Lorenz relation, whose dilute limit is tied to forward scattering, the optical theorem and Rayleigh scattering. In the same picture, reflection and refraction are produced by the polarization of the whole volume, with an additional, genuinely surface contribution for p polarization, distinct in origin from the local-field correction. A small coupled-dipole simulation confirms the self-consistent local-field relation, and the difficulties it meets at higher density are themselves a manifestation of the local-field enhancement. The everyday fact of refraction is thus traced back, with elementary means, to the atomic nature of matter.

\section*{Acknowledgments}
G.R.\ thanks his former student, colleague and friend Antoniangelo Agnesi for fruitful discussions, and Claude (Anthropic) for assistance with the numerical analysis and the preparation of the manuscript.

\appendix
\section{The integral of Eq.~(\ref{eq:sphere})}
\label{app:integral}

\begin{figure}[!htbp]
\centering
\includegraphics[width=0.8\textwidth]{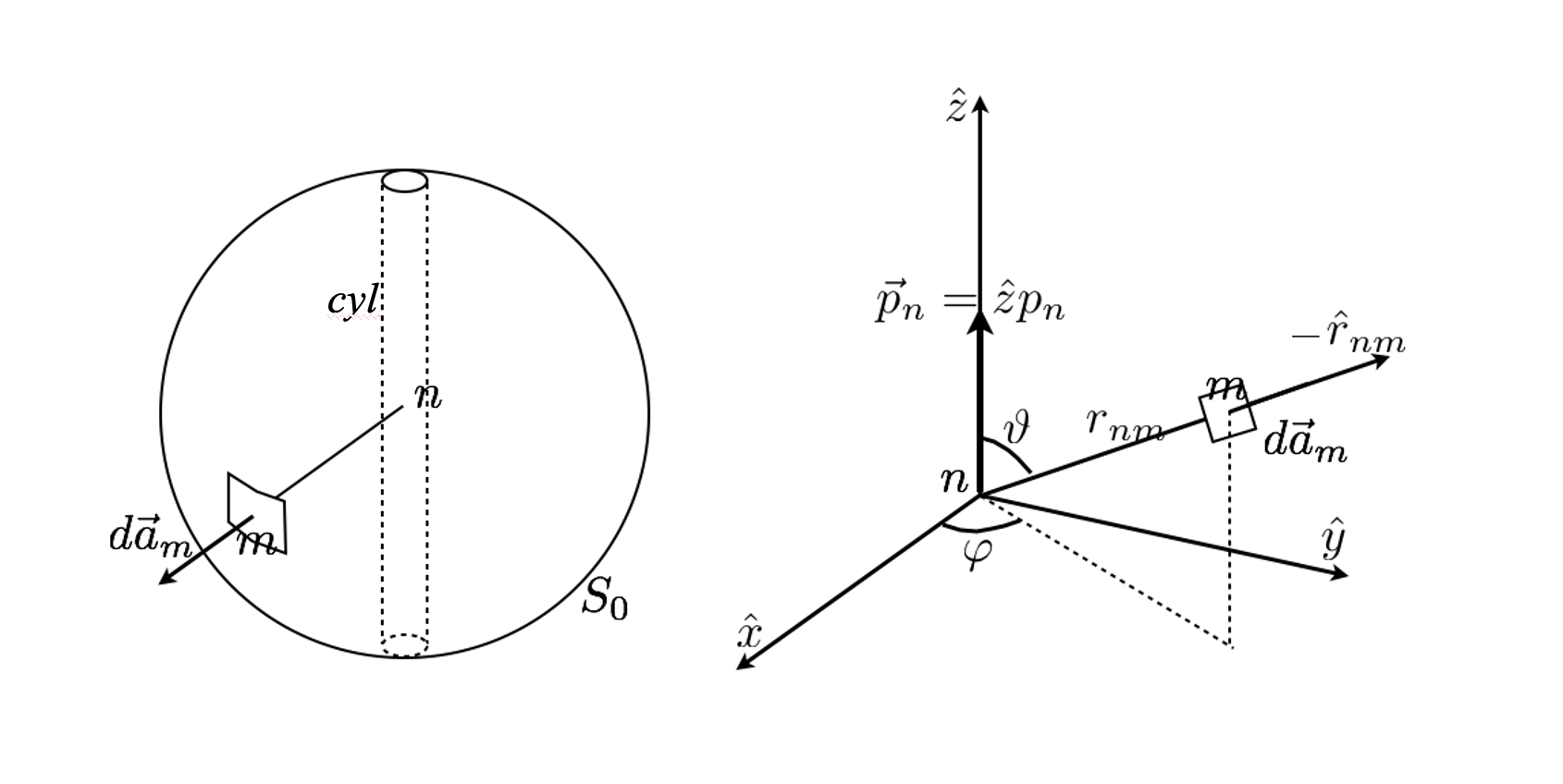}
\caption{Geometry of the integral of Appendix~\ref{app:integral}. Left: the sphere $S_0$ centred on unit $n$, the excluded needle and the outward area element $d\vv a_m$. Right: polar coordinates with $\vv p_n=\hat z p_n$; the outward unit vector is $\hat r_{mn}=-\hat r_{nm}$.}
\label{fig:geometry}
\end{figure}

For a constant vector $\vv A$ the vector identities for $\nabla\times(\vv A\times\vv B)$ and $\nabla(\vv A\cdot\vv B)$ combine into
\begin{equation*}
\nabla\times(\vv A\times\vv B)=-\nabla(\vv A\cdot\vv B)+\vv A\times(\nabla\times\vv B)+\vv A(\nabla\cdot\vv B).
\end{equation*}
In the quasi-static field of Eq.~\eqref{eq:quasistatic}, $\nabla_n=-\nabla_m$ acting on functions of $\vv r_{nm}$, and $\vv p_m\simeq\vv p_n$; with $\vv A=\vv p_n$ and $\vv B=\vv r_{nm}/r_{nm}^3$,
\begin{equation}
4\pi\eps\,\vv E_{\mathrm{rad},m}(\vv r_n)=\nabla_m\!\left(\frac{\vv p_n\cdot\vv r_{nm}}{r_{nm}^3}\right)-\vv p_n\times\left(\nabla_m\times\frac{\vv r_{nm}}{r_{nm}^3}\right)-\vv p_n\left(\nabla_m\cdot\frac{\vv r_{nm}}{r_{nm}^3}\right).
\label{eq:A1}
\end{equation}
The second term vanishes because $\vv r/r^3$ is irrotational; the third because the divergence of $\vv r_{nm}/r_{nm}^3$ is proportional to $\delta(\vv r_{nm})$, which vanishes in the integration domain, since this excludes $\vv r_n$. The first term becomes a surface integral (Fig.~\ref{fig:geometry}). On the lateral surface of the needle, $d\vv a_m$ is perpendicular to $\vv p_n$ and the contribution vanishes as the needle becomes thin. On the sphere, with $\vv p_n=p_n\hat z$ and $d\vv a_m=-\hat r_{nm}r_{nm}^2d\Omega_m$ (outward from $S_0$),
\begin{equation}
\int_{S_0}d\vv a_m\,\frac{\vv p_n\cdot\vv r_{nm}}{r_{nm}^3}=-p_n\int d\Omega\,\hat r(\hat r\cdot\hat z)=-\frac{4\pi}{3}\vv p_n,
\end{equation}
and division by $4\pi\eps$ gives Eq.~\eqref{eq:sphere}.

\section{Depolarization factors: three elementary cases}
\label{app:depol}

\begin{figure}[!htbp]
\centering
\includegraphics[width=0.85\textwidth]{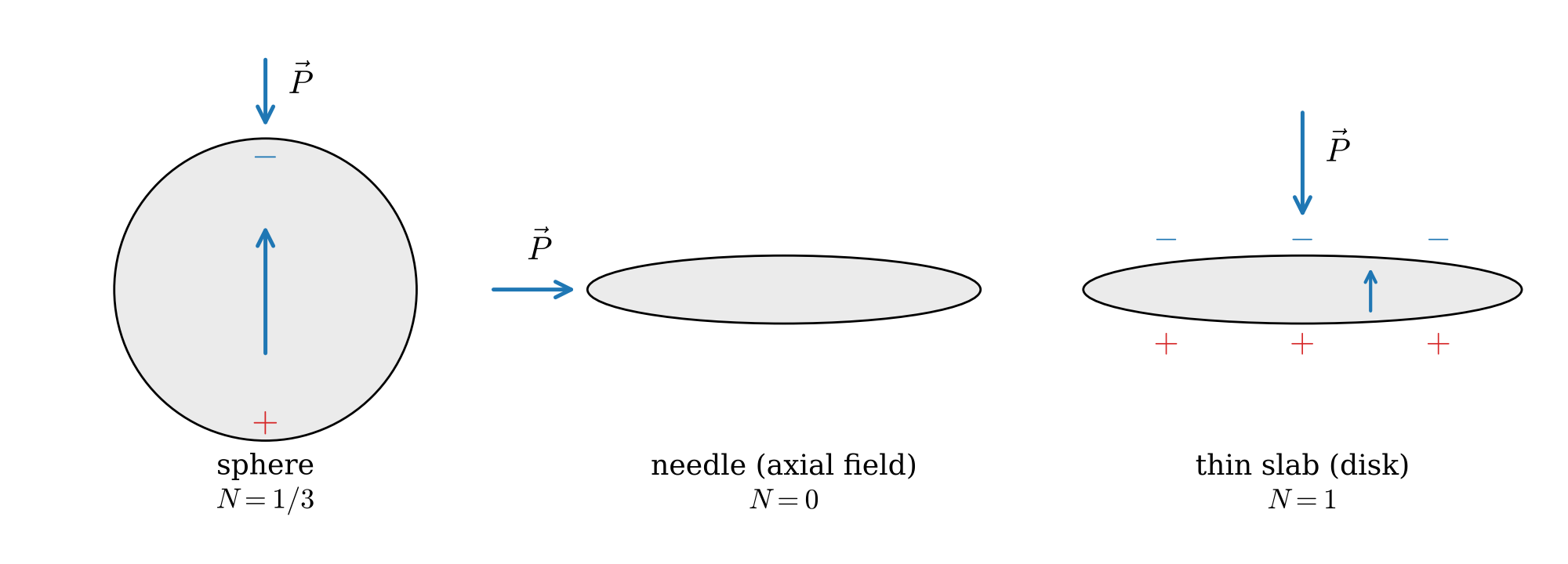}
\caption{Bound charges $\sigma_b=\vv P\cdot\hat n$ and depolarization field (inner arrow) for a sphere, a needle polarized along its axis, and a thin slab polarized normally to its faces.}
\label{fig:depol}
\end{figure}

The field inside a uniformly polarized ellipsoid is uniform~\cite{stratton}, $\vv E_{\mathrm{dep}}=-\mathsf N\cdot\vv P/\eps$, with $\mathsf N$ diagonal along the principal axes and $N_a+N_b+N_c=1$. The general $N_i$ are elliptic integrals~\cite{osborn,stoner}, but the three limits of Fig.~\ref{fig:depol} are elementary~\cite{portis}.

\noindent\emph{Sphere, $N=1/3$.} A uniformly polarized sphere is two uniformly charged spheres, of densities $\pm\rho_q$, displaced by $\vv d$, with $\vv P=\rho_q\vv d$. The field inside each is $\pm\rho_q\vv r/3\eps$, so their sum is the uniform field $-\vv P/3\eps$.

\noindent\emph{Needle along $\vv P$, $N=0$.} On the lateral surface $\hat n\perp\vv P$, so $\sigma_b=0$ and there is no depolarization field.

\noindent\emph{Slab normal to $\vv P$, $N=1$.} The faces carry $\sigma_b=\pm P$, as the plates of a capacitor, and the field between them is $-\vv P/\eps$.

The needle is the exclusion used to define the macroscopic field, Eq.~\eqref{eq:Emacro}; it brings no depolarization of its own, so the whole local-field correction, Eq.~\eqref{eq:Elocal}, comes from the sphere $S_0$. The slab shows the same mathematics at work at a real interface, where it produces the macroscopic bound-charge term of Section~\ref{sec:reflection}: same technique, different physics.


\begin{thebibliography}{99}

\bibitem{born_wolf} M. Born and E. Wolf, \emph{Principles of Optics}, 7th ed. (Cambridge University Press, 1999), Sec.~2.4.

\bibitem{reali1982} G. C. Reali, ``Reflection from dielectric materials,'' Am. J. Phys. \textbf{50}, 1133--1136 (1982).

\bibitem{fearn} H. Fearn, D. F. V. James, and P. W. Milonni, ``Microscopic approach to reflection, transmission, and the Ewald--Oseen extinction theorem,'' Am. J. Phys. \textbf{64}, 986--995 (1996).

\bibitem{hagenvik} H. O. H{\aa}genvik, K. Bl{\o}tekj{\ae}r, and J. Skaar, ``Dielectric media considered as vacuum with sources,'' Am. J. Phys. \textbf{85}, 830--833 (2017).

\bibitem{lai_lau_wong} H. M. Lai, Y. P. Lau, and W. H. Wong, ``Understanding wave characteristics via linear superposition of retarded fields,'' Am. J. Phys. \textbf{70}, 173--179 (2002).

\bibitem{ballenegger} V. C. Ballenegger and T. A. Weber, ``The Ewald--Oseen extinction theorem and extinction lengths,'' Am. J. Phys. \textbf{67}, 599--605 (1999).

\bibitem{berman} D. H. Berman, ``An extinction theorem for electromagnetic waves in a point dipole model,'' Am. J. Phys. \textbf{71}, 917--924 (2003).

\bibitem{jakoby} B. Jakoby and K. Hingerl, ``Do electromagnetic waves (such as light) really slow down in dielectrics?,'' Eur. J. Phys. \textbf{47}, 035203 (2026).

\bibitem{feynman_31} R. P. Feynman, R. B. Leighton, and M. Sands, \emph{The Feynman Lectures on Physics}, Vol.~1 (Addison-Wesley, 1963), Chs.~30-7 and 31.

\bibitem{feynman_33} R. P. Feynman, R. B. Leighton, and M. Sands, \emph{The Feynman Lectures on Physics}, Vol.~1 (Addison-Wesley, 1963), Sec.~33-6.

\bibitem{reali2014} G. Reali, ``A note on Feynman's calculation of reflection amplitudes for radiation striking a glass surface,'' Eur. J. Phys. \textbf{35}, 045022 (2014).

\bibitem{goldberger_watson} M. L. Goldberger and K. M. Watson, \emph{Collision Theory} (Wiley, 1964), Ch.~11, pp.~772--775.

\bibitem{jackson} J. D. Jackson, \emph{Classical Electrodynamics}, 3rd ed. (Wiley, 1999), Sec.~9.2 and Ch.~16.

\bibitem{nussenzveig} H. M. Nussenzveig, \emph{Causality and Dispersion Relations} (Academic Press, 1972).

\bibitem{purcell_pennypacker} E. M. Purcell and C. R. Pennypacker, ``Scattering and absorption of light by nonspherical dielectric grains,'' Astrophys. J. \textbf{186}, 705--714 (1973).

\bibitem{draine_flatau} B. T. Draine and P. J. Flatau, ``Discrete-dipole approximation for scattering calculations,'' J. Opt. Soc. Am. A \textbf{11}, 1491--1499 (1994).

\bibitem{stratton} J. A. Stratton, \emph{Electromagnetic Theory} (McGraw-Hill, 1941), Sec.~3.27.

\bibitem{osborn} J. A. Osborn, ``Demagnetizing factors of the general ellipsoid,'' Phys. Rev. \textbf{67}, 351--357 (1945).

\bibitem{stoner} E. C. Stoner, ``The demagnetizing factors for ellipsoids,'' Philos. Mag. \textbf{36}, 803--821 (1945).

\bibitem{portis} A. M. Portis, \emph{Electromagnetic Fields} (Wiley, 1978), pp.~135--140.

\end{thebibliography}
\end{document}